\documentclass[a4paper,reqno]{amsart}

\usepackage{amssymb}
\usepackage{srcltx}
\usepackage{graphicx}

\newcommand{\mathds}{\mathbb}

\theoremstyle{plain}
\newtheorem{theorem}{Theorem}[section]

\theoremstyle{remark}
\newtheorem{remark}{Remark}[section]

\theoremstyle{definition}

\numberwithin{equation}{section}

\def\Om{\Omega}
\def\om{\omega}
\def\e{\varepsilon}
\def\g{\gamma}

\def\l{\lambda}
\def\p{\partial}
\def\D{\Delta}

\def\E{\mbox{\rm e}}
\def\a{\alpha}
\def\b{\beta}

\def\L{\Lambda}

\def\vp{\varphi}

\def\Odr{\mathcal{O}}
\def\H{W_2}

\def\Hinf{W_\infty}
\def\Ho{\mathring{W}_2}
\def\Hoper{\mathring{W}_{2,per}}

\def\di{\,\mathrm{d}}

\def\iu{\mathrm{i}}

\def\la{\langle}
\def\ra{\rangle}

\def\Op{\mathcal{H}}

\def\cB{\mathcal{B}}

\DeclareMathOperator{\spec}{\sigma}

\DeclareMathOperator{\dist}{dist}

\newcommand{\RR}{\mathds{R}}
\newcommand{\ZZ}{\mathds{Z}}
\newcommand{\NN}{\mathds{N}}

\begin{document}

\title[Periodic waveguides]{Quantum waveguides with small periodic perturbations: gaps and edges of Brillouin zones}

\author{Denis Borisov}

\address{Institute of Mathematics of Ufa Scientific Center of Russian Academy of Sciences and
Bashkir State Pedagogical University, Ufa, Russia}

\author{Konstantin Pankrashkin}

\address{Laboratoire de m\'athematiques (UMR 8628), Universit\'e Paris-Sud 11, Orsay, France}

\begin{abstract}
We consider small perturbations of the Laplace operator in a multi-dimensional cylindrical domain by second order differential operators with periodic coefficients. We show that under certain non-degeneracy conditions such perturbations can open a gap in the continuous spectrum
and give the leading asymptotic terms for the gap edges. We also estimate the values of quasi-momentum at which the spectrum edges are attained. The general machinery is illustrated by several new examples in two- and three-dimensional structures.
\end{abstract}

\maketitle

\section{Introduction}

This paper is devoted to the spectral analysis of some periodic elliptic differential operators in unbounded periodic domains.
The spectrum of such an operator typically represents a half-line from which one removes a family of open intervals called \emph{gaps};
an important point is that this family can be empty, and it is one of the central questions to understand whether
a given operator really has a gap in its spectrum~\cite{kuch}. The importance of such questions is motivated by various applications.
For example, the operators of the above type appear in the study of photonic crystals:
the coefficients of the operator and the domain in which it acts describe the properties of the material,
and the gaps correspond to the energies at which no photons can be transmitted through the
sample \cite{PK}.

The existence of gaps for various periodic operators attracted a lot of attention in the last decade,
see e.g. the reviews \cite{HP,LP}, and we mention, in particular, the recent papers \cite{cnp,naz2,naz1}
discussing the opening of gaps for tube-like domains with various perturbations, but
our immediate interest arised from the papers \cite{HKSW,EKW} discussing some specific aspects
of the periodic spectral theory.
Recall that if an operator is periodic in $n$ directions, then
with the help of the Floquet-Bloch theory one obtains its spectrum
as the union of the ranges of the so-called \emph{band functions} $\tau\mapsto E_j(\tau)$, $j\in\NN$,
where the \emph{quasi-momentum} $\tau$ runs through a certain set $\cB\subset \RR^n$
called the \emph{Brillouin zone}; the functions $E_j$ are obtained as the eigenvalues of some
$\tau$-dependent quasiperiodic eigenvalue problem (see below for the precise definitions).
The authors of \cite{HKSW} discussed the position of the extrema of the band functions (\emph{spectral edges})
inside the Brillouin zone for generic periodic operators, and the paper \cite{EKW}
was devoted specifically to the operators with the one-dimensional periodicity.
In that case one has simply $\cB=(-\pi/T,\pi/T)$, where $T$ is the period, and it
was shown that the spectral edges are generically attained at values $\tau$ different from the symmetric
points $\pm\pi/T$ and $0$ corresponding to the energies at which one has either $T$-periodic or $T$-anti-periodic eigenfunction.
This observation should be 
viewed as a precaution: it is well known that the knowledge of the periodic and anti-periodic
eigenvalues is sufficient to describe the spectrum of one-dimensional periodic Schr\"odinger operators \cite{east},
which is the most studied periodic operator,
but this should not be extended by analogy to more general operators as it was sometimes assumed by mistake, see the respective
references in \cite{HKSW}. We also mention the recent paper \cite{CAMG} studying
the spectral edges for a particular problem of the solid state physics.

One of the most natural classes of one-periodic systems is provided by the waveguides,
but it seems that there are just few works discussing the shape of the band functions.
The paper \cite{acoust} discussed the position of the spectral edges
for acoustic structures, and our papers \cite{bp1,bp2} studied the problem
for a configuration consisting of two interacting waveguides
modeled by the free Laplacian,
and this discussion was continued in the very recent work \cite{naz3}
at the example of a cylinder with a periodic system of holes.
%
%discussing related questions
%at the physical level of rigor.

In the present paper we prove a general result giving sufficient conditions for the gap opening
for a class of second-order differential operators in multi-dimensional cylindrical domains.
It is shown that the presence of gaps at the energies different from the periodic and anti-periodic eigenvalues
is a generic fact, and we discuss the parameters controlling the gap opening
at various values of the quasi-momentum. Like in \cite{bp1,bp2,naz3},
the gap opening may occur at the points where the band functions of the reference operator
(the free Laplacian) meet in some specific way, but
the further analysis is done with the help of the elementary tools of the regular first-order perturbation
theory, while the previous works used the matched asymptotic expansions or other advanced methods
which are rather sensitive to the type of perturbation. We discuss in detail several new examples and show
that the gaps at the interior points of the Brillouin zone and at its center
are controlled by different parameters. In particular, we put in evidence
several perturbations which open gaps at intermediate values of the quasimomentum,
but the leading asymptotic term appears to vanish at the center of the Brillouin zone.
Therefore, our analysis considerably extends the class of situations for which
one can prove the existence of gaps compared to the previous works.

\section{Formulation of the problem and the main result}

\begin{figure}
\begin{center}\includegraphics[height=40mm]{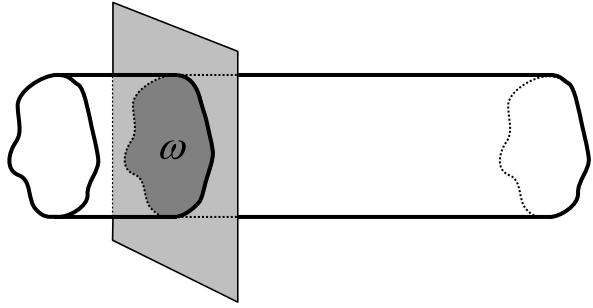}
\caption{A finite piece of the infinite cylinder $\Om$ with the cross-section $\om$.\label{fig1}}
\end{center}
\end{figure}

Let $x'=(x_1,\ldots,x_{n-1})$, $x=(x',x_n)$ be Cartesian coordinates in $\mathds{R}^{n-1}$ and $\mathds{R}^n$, respectively, $n\geqslant 2$,
and $\om$ be an open %
connected bounded domain in $\mathds{R}^{n-1}$ with Lipschitz boundary. By $\Om$ we denote an infinite straight cylinder in $\mathds{R}^n$ with the base  $\om$, namely, $\Om:=\om\times\mathds{R}$ (cf. Fig.~\ref{fig1}), and by $\Op_0$ we denote the positive Dirichlet Laplacian in $L_2(\Om)$ on the domain  $\Ho^2(\Om)$, which is introduced as the subspace of the functions in $\H^2(\Om)$ with zero trace on $\p\Om$. We assume that
the boundary $\p\om$ is sufficiently regular for the operator $\Op_0$ to be self-adjoint in $L_2(\Om)$.
Introduce another differential operator $\mathcal{L}_\e$ with the domain $\Ho^2(\Om)$
whose coefficients may depend on a small positive parameter $\e$,
\begin{equation*}
\mathcal{L_\e}:=-\sum\limits_{i,j=1}^{n} \frac{\p}{\p x_i} A_{ij} \frac{\p}{\p x_j} + \iu\sum\limits_{j=1}^{n} \left( A_j \frac{\p}{\p x_j}+\frac{\p}{\p x_j} A_j \right) + A_0,
\end{equation*}
where $A_{ij}\in \Hinf^1(\Om)$ are complex-valued functions satisfying $A_{ij}=\overline{A_{ji}}$, and  $A_j\in \Hinf^1(\Om)$,  $A_0\in L_\infty(\Om)$ are
real-valued functions. All the coefficients are supposed to be $T$-periodic in $x_n$:
\begin{align*}
A_{ij}(x',x_n+T,\e)&=A_{ij}(x,\e),& i,j&=1,\dots,n,\\
A_j(x',x_n+T,\e)&=A_j(x,\e),& j&=0,\dots, n,
\end{align*}
with an $\e$-independent period $T$. We assume that for $\e\to+0$ one has
\begin{equation}\label{1.2a}
\begin{aligned}
\sum\limits_{i,j=1}^{n} & \big\|A_{ij}(\cdot,\e)-A_{ij}(\cdot,0)\big\|_{\Hinf^1(\square)}
+\sum\limits_{j=1}^{n}
\big\|A_j(\cdot,\e)-A_j(\cdot,0)\big\|_{\Hinf^1(\square)}
\\
&+
\big\|A_0(\cdot,\e)-A_0(\cdot,0)\big\|_{L_\infty(\square)}=:\eta(\e)\to+0,
\end{aligned}
\end{equation}
where $\square:=\om\times(0,T)$ is the elementary cell. The condition  \eqref{1.2a} is satisfied, if, for example, all the coefficients are continuously differentiable  in both $x$ and $\e$.

Under the above assumptions and for sufficiently small
$\e$ the operator $\mathcal{L_\e}$ is symmetric and relatively bounded w.r.t.  $\Op_0$, and, by Kato-Rellich theorem,
for small $\e$ one can define the sum $\Op_\e:=\Op_0+\e\mathcal{L_\e}$ which is a self-adjoint operator in   $L_2(\Om)$.
One can consider $\Op_\e$ as the Hamiltonian of a quantum particle in the waveguide $\Om$.

Due to the periodicity, the operator $\Op_\e$ can be studied using the Floquet-Bloch theory
\cite{kuch}. Introduce self-adjoint operators
\begin{align}
\Op_\e(\tau)&:=-\D_{x'}+%\left(\iu\frac{\p}{\p %x_n}-\tau\right)^2
+l_n^2(\tau)
+\e \mathcal{L}_\e(\tau),\label{2.3}
\\
\mathcal{L_\e}(\tau)&:= \sum\limits_{i,j=1}^{n} l_i(\tau) A_{ij} l_j(\tau) + \sum\limits_{j=1}^{n} \big(A_j l_j(\tau)+ l_j(\tau) A_j \big) + A_0,\label{2.4}
\\
l_j(\tau)&:=\iu\frac{\p}{\p x_j},\quad j=1,\ldots,n-1,\quad l_n(\tau):=\iu\frac{\p}{\p x_n}-\tau,\nonumber
\end{align}
depending on the parameter
$\tau\in\cB=(-\pi/T,\pi/T]$ and acting in  $L_2(\square)$ on the domain
$\Hoper^2(\square)$ consisting of the functions from  $\H^2(\square)$ with zero trace on  $\p\om\times(0,T)$ and satisfying periodic boundary condition at   $x_n=0$ and $x_n=T$. The parameter $\tau$ is referred to as the \emph{quasi-momentum} and the set $\cB$ is usually called the \emph{Brillouin zone}.

For any $\tau$, the operator $\Op_\e(\tau)$ has a 
compact resolvent, and its spectrum consists of an infinite sequence of discrete real eigenvalues $E_m(\tau,\e)$, $m\geqslant 1$, which we assume to be ordered in the ascending order counting multiplicity. The mappings   $\tau\mapsto E_m(\tau,\e)$ are called \emph{band functions}; it is known that they are continuous with $E_m(-\pi/T,\e)=E_m(\pi/T,\e)$ and that the spectrum of $\Op_\e$
is the union of their ranges,
\begin{equation*}
\spec(\Op_\e)=\bigcup\limits_{m=1}^{\infty} \Big\{E_m(\tau,\e): \tau\in\cB\Big\}.
\end{equation*}
In particular,
$\inf \sigma(\Op_\e)=\Sigma_\e=\min_{\tau\in \cB} E_1(\tau,\e)$.
Each open interval $(\alpha,\beta)\subset [\Sigma_\e,+\infty)\setminus \sigma(\Op_\e)$ with $\alpha,\beta \in\sigma(\Op_\e)$ is called a \emph{gap} of $\Op_\e$. We note that the unperturbed operator  $\Op_0$ has no gaps and its spectrum fills a semi-axis (see below). Note that under some additional assumptions one can show that the band functions are not constant on any interval \cite{bent,fried,fil,sobol};
%KP
this property is not used in our constructions.
The main aim of the present paper is to obtain some conditions guaranteeing the existence of gaps for $\Op_\e$ and
to determine the values of the quasi-momentum  $\tau$ at which the band functions attain the respective gap edges $\a$ and $\b$.

The eigenvalues of $\Op_0(\tau)$ can be found by the separation of variables. Denote by $-\D^{(D)}_{\om}$  the positive Dirichlet Laplacian in
$\om$. Since this operator is self-adjoint and has a compact resolvent, its spectrum consists of real eigenvalues of finite multiplicity
denoted by  $\l_j$, $j=1,2,\dots$, and assumed to be ordered in the ascending order counting multiplicity.
The spectrum of $\Op_0(\tau)$ then consists of the eigenvalues
\begin{equation}\label{2.6}
\L_{j,p}(\tau)=\l_j+\left(\tau+\frac{2\pi p}{T}\right)^2,\quad j\in\NN, \quad p\in\ZZ,
\end{equation}
and the values $E_m(\tau,0)$ are obtained by their rearrangement in the ascending order.
Moreover, if   $\psi_j(x')$ are the eigenfunctions of $-\D^{(D)}_{\om}$ associated with the eigenvalues   $\l_j$
and chosen real and orthonormal, then the eigenfunctions $\Psi_{j,p}$ of $\Op_0(\tau)$ for the eigenvalues   $\L_{j,p}(\tau)$
can be written as
\begin{equation}\label{2.7}
\Psi_{j,p}(x):=\frac{1}{T^{1/2}} \psi_j(x')\E^{\frac{2\pi \iu p}{T}x_n},
\end{equation}
and they are orthonormal in $L_2(\square)$.

Given two functions $\L_{j,p}$ and $\L_{k,q}$ defined by (\ref{2.6}), we let
%Define the functions
\begin{align}
\b_\pm(\tau):=&
\pm\frac{|b_{12}^0(\tau)|}{|k_3(\tau)|}\sqrt{k_3^2(\tau)-k_1^2(\tau)}-\frac{k_1(\tau) k_4(\tau)}{k_3(\tau)}+k_2(\tau),
\label{2.12}
\end{align}
where
\begin{align}
&
\begin{aligned}
k_1(\tau)&:=-\frac{\pi(p+q)}{T}-\tau, & k_2(\tau)&:=-\frac{b_{11}^0(\tau)+b_{22}^0(\tau)}{2},\\
k_3(\tau)&:= \frac{\pi}{T}(p-q),& k_4(\tau)&:=\frac{b_{22}^0(\tau)-b_{11}^0(\tau)}{2}.
\end{aligned}\label{2.18}
\\
&B_0(\tau):=
\begin{pmatrix}
b^0_{11}(\tau)  & b^0_{12}(\tau)
\\
b^0_{21}(\tau) & b^0_{22}(\tau)
\end{pmatrix}=
\begin{pmatrix}
\big\la\mathcal{L}_0(\tau)\Psi_{j,p},\Psi_{j,p}\big\ra_{L_2(\square)}  &\big\la\mathcal{L}_0(\tau)\Psi_{j,p},\Psi_{k,q}\big\ra_{L_2(\square)}
\\
\big\la\mathcal{L}_0(\tau)\Psi_{k,q},\Psi_{j,p}\big\ra_{L_2(\square)}  &\big\la\mathcal{L}_0(\tau)\Psi_{k,q},\Psi_{k,q}\big\ra_{L_2(\square)}
\end{pmatrix},\label{3.14}
\end{align}
as $\tau\geqslant 0$, and
\begin{align}
&
\begin{aligned}
k_1(\tau)&:=\frac{\pi(p+q)}{T}-\tau, & k_2(\tau)&:=-\frac{b_{11}^0(\tau)+b_{22}^0(\tau)}{2},\\
k_3(\tau)&:= \frac{\pi}{T}(q-p),& k_4(\tau)&:=\frac{b_{22}^0(\tau)-b_{11}^0(\tau)}{2}.
\end{aligned}\label{2.18a}
\\
&B_0(\tau):=
\begin{pmatrix}
b^0_{11}(\tau)  & b^0_{12}(\tau)
\\
b^0_{21}(\tau) & b^0_{22}(\tau)
\end{pmatrix}\\
&\qquad\qquad=
\begin{pmatrix}
\big\la\mathcal{L}_0(\tau)\Psi_{j,-p},\Psi_{j,-p}\big\ra_{L_2(\square)}  &\big\la\mathcal{L}_0(\tau)\Psi_{j,-p},\Psi_{k,-q}\big\ra_{L_2(\square)}
\\
\big\la\mathcal{L}_0(\tau)\Psi_{k,-q},\Psi_{j,-p}\big\ra_{L_2(\square)}  &\big\la\mathcal{L}_0(\tau)\Psi_{k,-q},\Psi_{k,-q}\big\ra_{L_2(\square)}
\end{pmatrix},\label{3.14a}
\end{align}
as $\tau<0$.

 Let us formulate our main result.

\begin{theorem}\label{th2.1}
%Let
Suppose there exist numbers  $\tau_0\in\Big[0,\dfrac{\pi}{T}\Big]$, $j,k\in\{1,2\}$ and $p,q\in \mathds{Z}$ satisfying the following conditions:
\begin{gather}
\text{$\lambda_j$ and $\lambda_k$ are simple eigenvalues of  $-\D^{(D)}_{\om}$;}  \label{2.10b}
\\
\L_{j,p}(\tau_0)=\L_{k,q}(\tau_0)=:\L_0;\label{2.8}
\\
\frac{\p\L_{j,p}}{\p\tau}(\tau_0)\frac{\p\L_{k,q}}{\p\tau}(\tau_0)<0;\label{2.9}
\\
\text{$\L_0\notin \L_{l,s}\Big(\Big[0,\dfrac{\pi}{T}\Big]\Big)$ as $(l,s)\notin\{(j,p),(k,q)\}$}; \label{2.10a}
\\
\langle\mathcal{L}_0(\pm\tau_0)\Psi_{j,p},\Psi_{k,q}\rangle_{L_2(\square)}\not=0. \label{2.10}
\end{gather}
Then
\begin{enumerate} \def\theenumi{A\arabic{enumi}}
\item \label{A1}
The strict inequalities $\beta_-(\tau_0)<\beta_+(\tau_0)$
and $\beta_-(-\tau_0)<\beta_+(-\tau_0)$ hold true, %.
where the functions $\b_\pm$ are introduced by (\ref{2.12}) with aforementioned  $p$, $q$, $j$, and $k$. 
\item 
\label{A2} If, in addition, one has
\begin{equation}
\b_l:=\max\{\b_-(\tau_0),\b_-(-\tau_0)\}<
\b_r:=\min\{\b_+(\tau_0),\b_+(-\tau_0)\},\label{2.16}
\end{equation}
 then there exists  $\e_0>0$ such that for all $\e\in(0,\e_0)$ the operator $\Op_\e$ has a spectral
 gap $\big(\a_l(\e),\a_r(\e)\big)$ whose edges have the asymptotics
\begin{equation}
\a_{l/r}(\e)=\L_0+\e\b_{l/r}+\Odr\big(\e^2+\e\eta(\e)\big),\label{2.11}
\end{equation}
and the associated band functions  $E_{l/r}(\e,\tau)$ of $\Op_\e$ attain the gap edges $a_{l/r}(\e)$ at the points  $\tau_{l/r}(\e)$,
\begin{equation}\label{2.13}
\min\limits_{\tau} E_r(\e,\tau)=E_r\big(\e,\tau_r(\e)\big),
\quad \max\limits_{\tau} E_l(\e,\tau)=E_l\big(\e,\tau_l(\e)\big),
\end{equation}
for which the asymptotics
\begin{align}
&\tau_{l/r}(\e)=\tau_0+\e \g_{l/r}+\Odr(\e^{3/2}+\e\eta^{1/2})
\label{2.14}
\\
&
\begin{aligned}
&\g_l:= \frac{k_1(\tau_*) |b_{12}^0(\tau_*)|}{|k_3(\tau_*)| \sqrt{k_3^2(\tau_*)-k_1^2(\tau_*)}}-\frac{k_4(\tau_*)}{k_3(\tau_*)},
\\
&\g_r:=-\frac{k_1(\tau_*) |b_{12}^0(\tau_*)|}{|k_3(\tau_*)| \sqrt{k_3^2(\tau_*)-k_1^2(\tau_*)}}-\frac{k_4(\tau_*)}{k_3(\tau_*)},
\end{aligned}\nonumber
\end{align}
are valid. In each of the latter formulas the number $\tau_*$ should be chosen to that of the values $\pm \tau_0$, at which the maximum or minimum is attained in the formulas (\ref{2.16}).

\item \label{A3}
If at least one the following two conditions is valid:
\begin{itemize}
\item
all the coefficients $\Op_\e$ are real
\item $\tau_0=0$,
\end{itemize}
then the condition  (\ref{2.16}) is satisfied.

\end{enumerate}
\end{theorem}

Before proceeding to the proof, let us give some comments on the assumptions and the statement of the theorem.

\begin{figure}
\centering

\begin{tabular}{ccc}
\includegraphics[height=40mm]{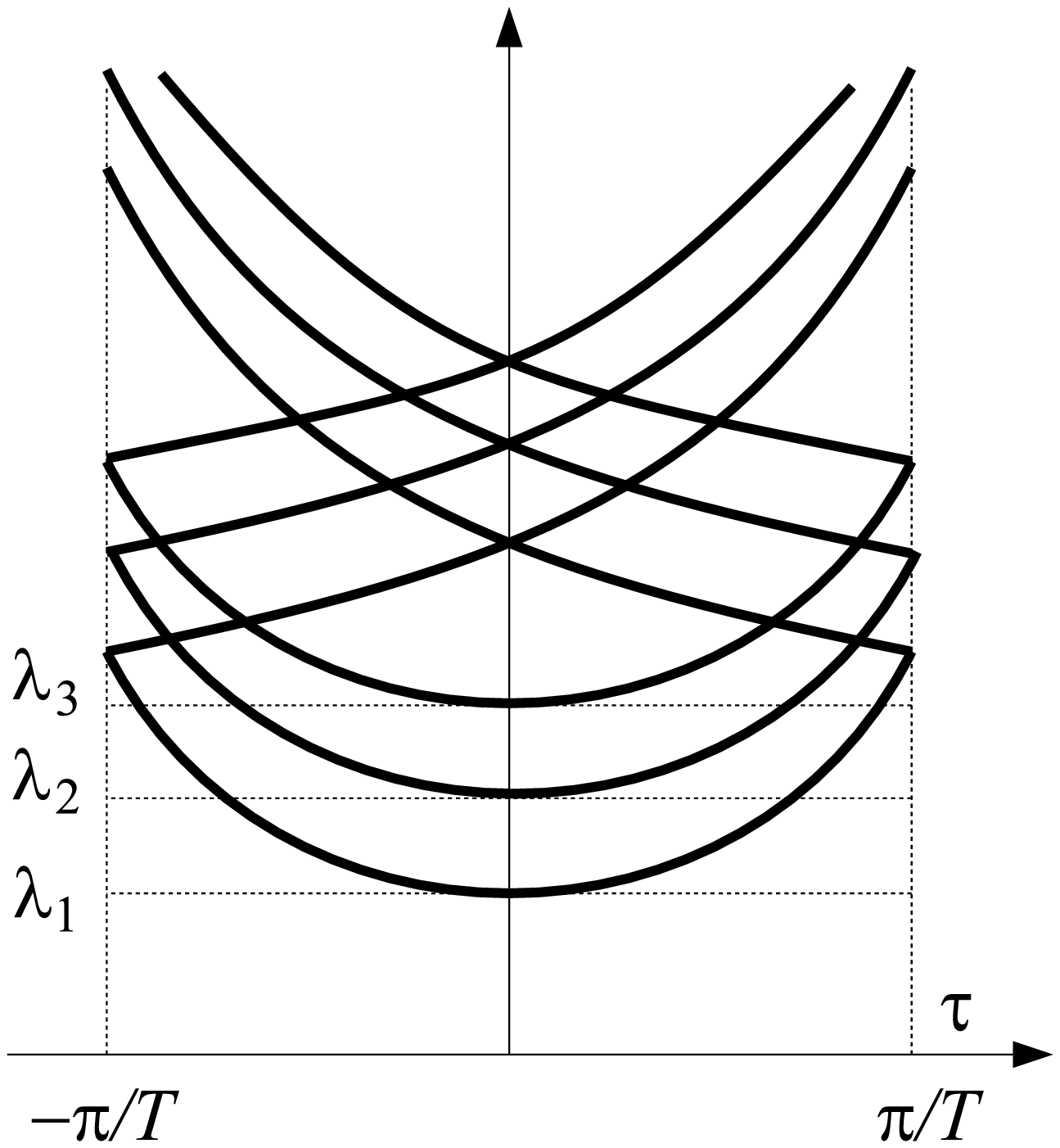}
&
\includegraphics[height=40mm]{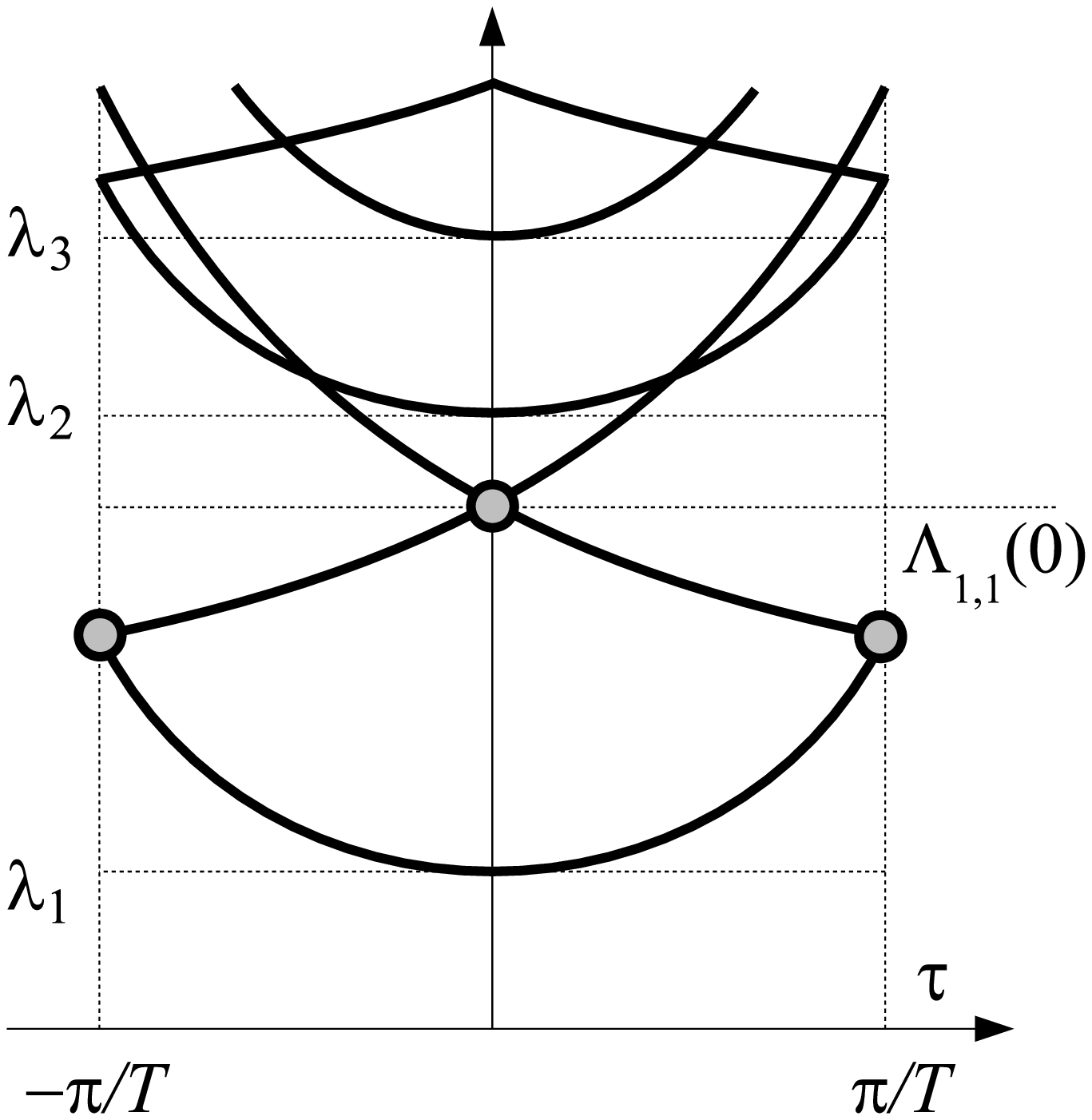}
&
\includegraphics[height=40mm]{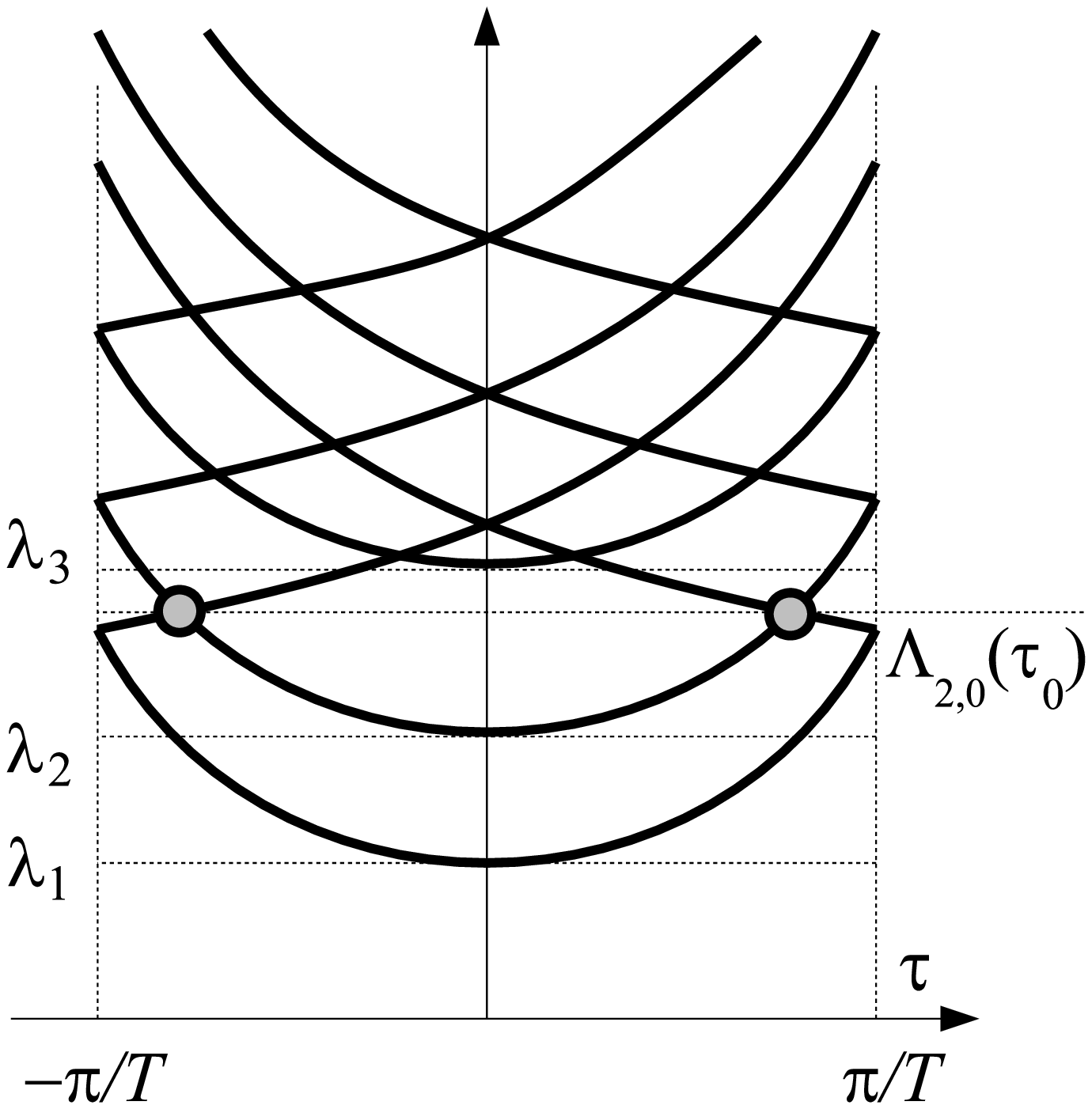}\\
(a) & (b) & (c)
\end{tabular}
\caption{Some possible configurations of the unperturbed band functions. The intersections satisfying the assumptions \eqref{2.8}, \eqref{2.9} and \eqref{2.10a}
are marked by small circles
\label{fig3}}
\end{figure}

\begin{remark}
The non-degeneracy condition (\ref{2.10b}) is independent of the period $T$ and concerns only the eigenvalues of the cross-section
operator $-\D^{(D)}_{\om}$. This condition allows us to reduce the spectral study of  $\Op_\e$ in a vicinity of $\L_0$ to
an eigenvalue splitting problem for a $2\times 2$ matrix. Note that this condition is always valid for $j=k=1$, because
the lowest eigenvalue of the Dirichlet Laplacian in a bounded connected domain is always non-degenerate.
\end{remark}

\begin{remark}\label{rm2.2}
The conditions \eqref{2.8}, \eqref{2.9} and \eqref{2.10a} impose some restrictions on the behavior of the band functions $\L_{j,p}$ and $\L_{k,q}$
an the intersection point. The condition (\ref{2.8}) means exactly the presence of an intersection. The condition  (\ref{2.9}) shows that
one of the functions increases and the other one decreases at the intersection point. Finally, the third condition (\ref{2.10a}) expresses
the fact that the projection of the intersection point on the ordinate axis should not be overlapped by the projections of the graphs of the remaining
band functions $\L_{l,s}(\tau)$ as $\tau\in[0,\pi/T]$. We observe that %it is possible
such situation can happen
only as $j+k\leqslant 3$.

One can easily see that the validity of these assumptions is conditioned by the presence of certain relations between the transversal eigenvalues  $\lambda_j$ and the period $T$. One can easily find examples where there are points in which all these assumptions hold, see e.g. Fig.~\ref{fig3}(a).

We remark first that the conditions  %(\ref{2.6}) and
(\ref{2.8}), (\ref{2.10a}) can hold true only for $p$ and $q$ having opposite signs. Moreover, due to  \eqref{2.6} one has
\begin{equation}
\label{eq-l12}
\lambda_k-\lambda_j= \dfrac{2\pi (p-q)}{T}\Big(\dfrac{2\pi(p+q)}{T}+2\tau_0\Big).
\end{equation}
In particular, if all three conditions in question hold with $\tau_0=0$, then $j=k=1$ and $q=-p$,
and if all the conditions are valid for $\tau_0=\pi/T$, then automatically $j=k=1$ and $q=-p-1$.

To guarantee the presence of at least one combination $(j,p)$,
$(k,q)$ satisfying the above assumptions for $\tau_0=0$, it is sufficient to ask for the inequality $\L_{1,1}(0)<\L_{2,0}(0)$, which is equivalent
to $T>2\pi/\sqrt{\lambda_2-\lambda_1}$, see Fig.~\ref{fig3}(b); in this case all three conditions are valid for $\tau_0=0$, $j=k=1$, $p=1$, $q=-1$.
On the other hand, to satisfy the conditions (\ref{2.8}) and (\ref{2.9}) for some $\tau_0\in(0,\pi/T)$ it is sufficient to obey, for instance, the inequality $\L_{2,0}(0)<\L_{1,1}(0)$, then one can take $j=1$, $k=2$, $p=-1$ and $q=0$, and the condition (\ref{2.10a}) is equivalent to
$\L_{2,0}(\tau_0)<\Lambda_{3,0}(0)$, see Fig.~\ref{fig3}(c).
Rewriting these inequalities with the help of the explicit expressions (\ref{2.6}), one can easily see
that the three conditions (\ref{2.8}), (\ref{2.9}), (\ref{2.10a}) hold true with some $\tau_0\in(0,\pi/T)$, $j=1$, $k=2$, $p=-1$ and $q=0$, if the period $T$ obeys the estimates
\[
\dfrac{2\pi}{\sqrt{\lambda_3  -
\lambda_1}+ %-
\sqrt{\lambda_3-\lambda_2}}
<T<
\dfrac{2\pi}{\sqrt{\lambda_2-\lambda_1}}.
\]
With the help of the explicit formulas for $\L_{j,p}$ one can easily construct other sufficient conditions.
Note that for any fixed cross-section $\om$ one can satisfy the assumption
by an appropriate choice of the period $T$, and, moreover, one can obtain in this way
any prescribed finite number of intersections satisfying all the assumptions.
One is also able to choose parameters in such a way that the assumption
will be satisfied for any 
prescribed value of $\tau_0$.
\end{remark}

\begin{figure}
\centering
\includegraphics[width=60mm]{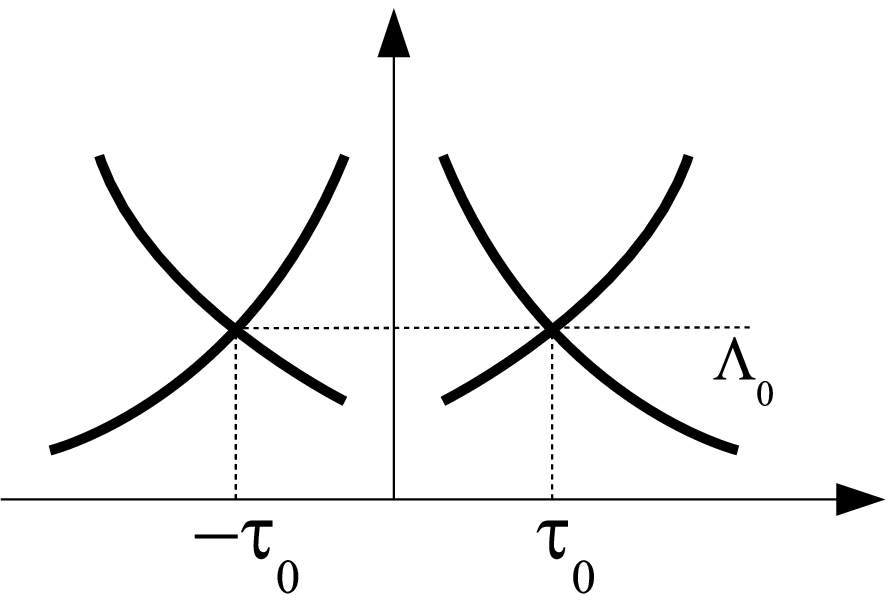}\\
(a) Intersection of the unperturbed band functions.\\
\includegraphics[width=80mm]{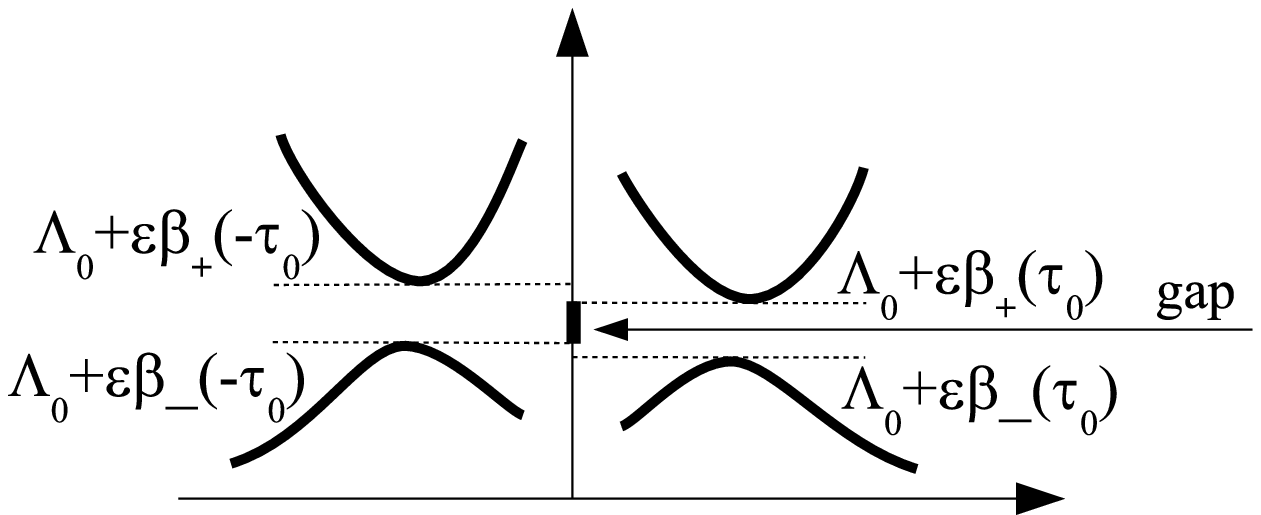}\\
(b) The assumption \eqref{2.16} is valid,
the perturbed band functions open a gap.\\
\includegraphics[width=80mm]{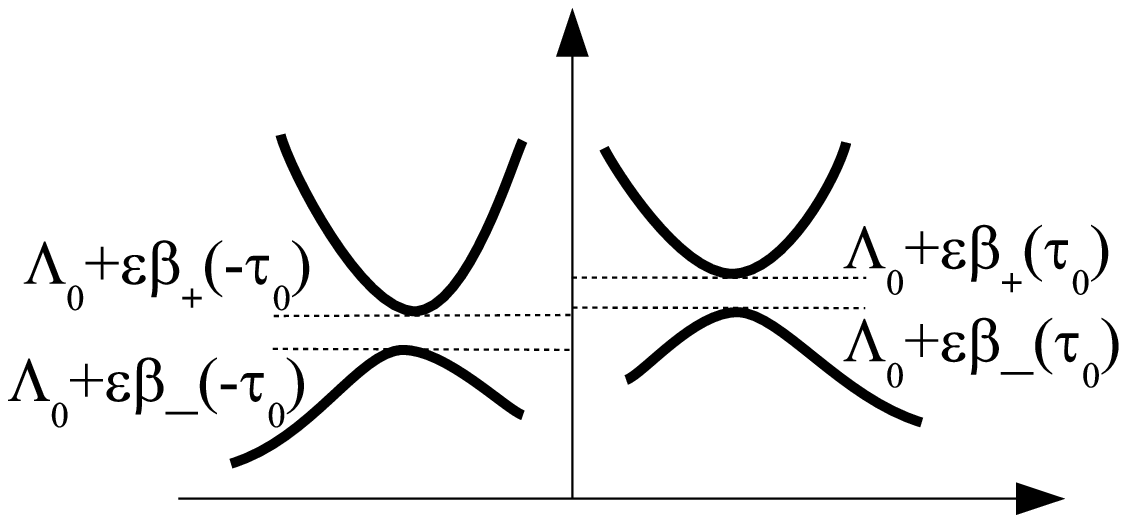}\\
(c) The assumption \eqref{2.16} is not satisfied, no gap opening
\caption{\label{fig4}}
\end{figure}

\begin{remark}
Let us explain the meaning of $\b_\pm$. As we shall show later, the perturbed operator $\mathcal{H}_\e$ has two band functions $E_\pm(\e,\tau)$ converging to $\L_{j,p}(\tau)$ and $\L_{k,q}(\tau)$ taken in an appropriate order. For each $\e$ the function $E_-(\e,\tau)$ a maximum in each of the segments $[-\pi/T,0]$ and $[0,\pi/T]$ (we exclude here the symmetric case $\tau_0=0$). The asymptotics of these maxima are $\L_0+\e \b_-(\pm \tau_0)+\Odr(\e^2+\e\eta)$ and this is why the global maximum of $E_-$ is exactly $\L_0+\e \b_l + \Odr(\e^2+\e\eta)$, see Fig.~\ref{fig4}. The function $\b_+$ appear in the same way and is associated with the minima of $E_+$.
\end{remark}

\begin{remark}
The key condition (\ref{2.10}) allows one to make some conclusions on the behavior of the perturbed band functions in a vicinity of  $\tau_0$
using some elementary tools of the regular perturbation theory. Checking this condition is usually the most difficult
step when applying the theorem, which will be seen with the series of examples below. As %
it follows from (\ref{A3}), for
operators with real-valued coefficients as well as  for $\tau_0=0$,
Eq. \eqref{2.10} is the only non-trivial assumption on the perturbation $\mathcal{L}_\e$.
\end{remark}

\begin{remark}%KP
If some of the coefficients of $\mathcal{L}_\e$ are non-real, then 
the quantities $\g_{l/r}$ appearing in (\ref{2.14}) may be associated with different values of $\tau_*$.
The operator $\mathcal{H}_\e$ is not real anymore, and its band functions
can be non-even w.r.t. $\tau$, and this is why their extremal points are not necessarily symmetric w.r.t. zero.
\end{remark}

\begin{remark}
The theorem can be extended to other types of unperturbed operators as far as  the unperturbed band functions have a similar structure. We do not develop this direction to avoid technicalities, but just remark that the results are valid in literally the same form for some other
boundary conditions  $\p\Om$, in particular, for the Neumann condition and the Robin condition with a constant coefficient,
and such a modification results in an appropriate redefinition of the eigenvalues $\lambda_j$ and the eigenfunctions $\psi_j$.
\end{remark}

\section{Proof of the main result}

This section is completely devoted to the proof of Theorem~\ref{th2.1}.
We do all the constructions for $\tau_0\in[0,\pi/T)$ only; the study of $\tau_0=\pi/T$ is completely identical, but it requires cumbersome notation since it corresponds to the end point of the Brillouin zone. % 
The proof consists of several main steps.

\subsection*{Step 1: Simplification of the perturbation} By (\ref{1.2a}) one can represent $\mathcal{L}_\e=\mathcal{L}_0+\eta(\e) \mathcal{M}_\e$, where $\mathcal{M}_\e$ is %the
an operator of the same type as $\mathcal{L}_\e$ except the fact that instead of the continuity in $\e$ at $\e=0$ we ask for the uniform boundedness for $\e<\e_1$ in the appropriate Sobolev norms. Under these assumptions and for sufficiently small $\e$
the operator  $\mathcal{M}_\e$ is relatively bounded w.r.t.  $\mathcal{K}_\e:=\Op_0 + \e \mathcal{L}_0$. By analogy with (\ref{2.4}) one can define the operators $\mathcal{M}_\e(\tau)$ and $\mathcal{K}_\e(\tau)$, and then $\mathcal{M}_\e(\tau)$ is also relatively bounded w.r.t. $\mathcal{K}_\e(\tau)$.

Denote by $\widetilde E_m(\tau,\e)$, $m=1,2,\dots$, the eigenvalues of  $\mathcal{K}_\e(\tau)$ taken in the ascending order counting multiplicity. By the minimax principle one can choose a constant $c>0$ such that for sufficiently small $\e$ the uniform in
$\tau\in\cB$ estimates
\begin{equation}
\label{3.1}
\begin{aligned}
\big|E_m(\tau,\e)-E_m(\tau,0)\big|&\leqslant c \e \big|E_m(\tau,0)\big|, \\
\big|\widetilde E_m(\tau,\e)-E_m(\tau,\e)\big|&\leqslant c\e\eta(\e) \big|E_m(\tau,0)\big|
\end{aligned}
\end{equation}
hold true. Since the required estimates in the assertion (\ref{A2}) of Theorem \ref{th2.1} are linear in $\e$, the second inequality in \eqref{3.1}
shows that it is sufficient to consider the case of an $\e$-independent perturbation, $\mathcal{L}_\e=\mathcal{L}_0$,
which will be assumed throughout the rest of the proof.

\subsection*{Step 2: Estimates for the shift of the band functions} At this step we obtain a general, rather rough estimate for the differences between the perturbed and limiting band functions. 
The explicit expressions show that for any $C>\Lambda_0$ there exists just a finite number of the pairs
$(l,s)\in\NN\times\ZZ$ with $(-\infty,C)\cap \Lambda_{l,s}(\cB)\ne\emptyset$. Pick an arbitrary $C_1>\Lambda_0$,
then, by  (\ref{3.1}), one can find $N\in \NN$ and $\e_2>0$ such that $E_m(\tau,\e)>C_1$
for all $m>N$, $\e<\e_2$ and $\tau\in\cB$. Thus, there exists  $C_2>0$ such that  $\big|E_m(\tau,\e)-E_m(\tau,0)\big|\leqslant C_2 \e$ for all $m\leqslant N$, $\e<\e_2$ and $\tau\in\cB$.

Now we find $M\in\NN$ for which $\Lambda_0=E_M(\tau_0,0)=E_{M+1}(\tau_0,0)$.
By the preceding estimates we can conclude that for any $\delta>0$
there exists $\e_3>0$ such that $(\Lambda_0-\delta,\Lambda_0+\delta)\cap E_j(\cB,\e)=\emptyset$
for all $j\notin\{M,M+1\}$ и $\e<\e_3$.
Thus, the spectrum of $\Op_\e$ near $\Lambda_0$ for small $\e$ is determined by the behavior
of the two band functions $E_M$ and $E_{M+1}$ near $\pm\tau_0$, and for $\e=0$ they coincide locally with the functions $\L_{j,p}$ and $\L_{k,q}$.

By (\ref{2.6}) we have $\L_{j,-p}(-\tau)=\L_{j,p}(\tau)$. This shows that the assumptions
 (\ref{2.8}), (\ref{2.9}) and (\ref{2.10}) are also valid with $j$, $p$, $k$, $q$, and $\tau_0$ replaced by $j$, $-p$, $k$, $-q$, and $-\tau_0$, respectively, and with the same value  $\L_0$. One can find a neighborhood of $\L_0$ having no intersections
 with the ranges of $\L_{l,s}$ as $(l,s)\notin\{(j,p),(k,q)\}$ and, moreover, it is clear that in $[-\pi/T,\pi/T]$ there are no other
 values $\pm\tau_0$ for which Eq. (\ref{2.8}) holds. To summarize, for $\e\in(0,\e_3)$ and for any $a>0$ there is
 a constant $t_0(a)>0$ such that
\begin{equation}\label{3.3}
\dist\Big(\spec\big(\Op_\e(\tau)\big),\, \L_0\Big)\geqslant a\,\e\quad \text{for}\quad |\tau\pm\tau_0|\geqslant t_0(a)\e.
\end{equation}
In other words, for the indicated values of $\tau$ the spectrum of $\Op_\e(\tau)$ is separated from $\L_0$ by a distance at least
$a\e$, and it is now sufficient to study the band functions of  $\Op_\e$ near $\L_0$ as $|\tau\pm\tau_0|\leqslant t_0(a)\e$.
We consider the case of $\tau_0$ only; the case of $-\tau_0$ is studied in the same way. This will be the subject of the next step.

\subsection*{Step 3: Asymptotics for the band functions in a vicinity of $\tau_0$} We let $\tau=\tau_0+\e\, t$, where $t\in[-t_0(a),t_0(a)]$ is a real-valued parameter. The number $\L_0$ is a double eigenvalue of $\Op_0(\tau_0)$, and the perturbation $\e \mathcal{L}_0(\tau_0+\e\, t)$ is regular, see (\ref{2.3}) and (\ref{2.4}), and we can apply the standard regular perturbation theory, see e.g. \cite[Ch. V\!I\!I, \S 3]{Kato}. The first of the estimates  \eqref{3.1} shows that for sufficiently small $\e_0>0$
the operator $\Op_\e(\tau_0+\e t)$ has two eigenvalues (counting multiplicity) that converge to  $\L_0$ as $\e\to+0$.
Denote these eigenvalues by $E_\pm(\e,\tau)$ and the associated eigenfunctions by  $\phi_\pm(x,\tau,\e)$. Since the perturbation by $\e\mathcal{L}_0(\tau_0+\e\,t)$ is regular and the operators   $\Op_\e(\tau_0+\e\,t)$  are self-adjoint, the eigenvalues $E_\pm(\e,\tau+\e t)$ and the associated eigenfunctions $\phi_\pm(x,\tau+\e t,\e)$ are holomorphic w.r.t. $\e$ (the latter are holomorphic
in $\Hoper^2(\square)$-norm), and the leading terms of their Taylor series near $\e=0$ are of the form:
\begin{gather}
E_\pm(\e,\tau_0+\e t)=\L_0+\e K_\pm(t)+\Odr(\e^2),\nonumber
\\
\phi_\pm(x,t,\e)=\Psi_\pm(x,t)+\e\Phi_\pm(x,t)+\Odr(\e^2), \label{3.4}
\\
\Psi_+:=c_1^+\Psi_{j,p}+c_2^+ \Psi_{k,q},\quad
\Psi_-:=c_1^-\Psi_{j,p}+c_2^-\Psi_{k,q},\nonumber
\end{gather}
where $c_i^\pm=c_i^\pm(t)$ are some constants  that do not vanish simultaneously. Since the eigenfunctions $\phi_\pm$ can be chosen orthonormal
in  $L_2(\square)$, the same is true for $\Psi_\pm$. This gives
\begin{equation}\label{3.7}
c_1^+\overline{c_1^-} + c_2^+ \overline{c_2^-}=0,\quad |c_1^\pm|^2+|c_2^\pm|^2=1,
\end{equation}
and the error estimates in (\ref{3.4}) are uniform for $t\in[-t_0(a),t_0(a)]$ for any fixed $a$.

To determine the coefficients $K_\pm$ it is sufficient to employ the regular perturbation theory. Namely, substitute the formulas  (\ref{3.4}) into the eigenvalue equation
\begin{equation*}
\Op_\e(\tau_0+\e t)\phi_\pm=E_\pm\phi_\pm
\end{equation*}
and equate the coefficients at the first power of  $\e$, then one arrives at the two equations
\begin{equation*}
\big(\Op_0(\tau_0)-\L_0\big)\Phi_\pm=2t l_n(\tau_0)\Psi_\pm + \mathcal{L}(\tau_0)\Psi_\pm+K_\pm\Psi_\pm.
\end{equation*}
These equations are solvable iff the functions at the right-hand side are orthogonal to  $\Psi_{j,p}$ and $\Psi_{k,q}$ in $L_2(\square)$:
\begin{equation}\label{3.6}
\begin{aligned}
\big\la2tl_n(\tau_0)\Psi_\pm+\mathcal{L}(\tau_0)\Psi_\pm +K_\pm\Psi_\pm,\Psi_{j,p}\big\ra_{L_2(\square)}&=0,
\\
\big\la 2tl_n(\tau_0)\Psi_\pm+\mathcal{L}(\tau_0)\Psi_\pm +K_\pm\Psi_\pm,\Psi_{k,q}\big\ra_{L_2(\square)}&=0.
\end{aligned}
\end{equation}
Using the orthonormality of $\Psi_{j,p}$ and $\Psi_{k,q}$ and Eqs. (\ref{2.7}) and (\ref{3.7})
we arrive at
\begin{gather*}
\big\la l_n(\tau_0)\Psi_{j,p},\Psi_{k,q}\big\ra_{L_2(\square)}= \big\la l_n(\tau_0)\Psi_{k,q},\Psi_{j,p}\big\ra_{L_2(\square)}=0,
\\
\big\la l_n(\tau_0)\Psi_{j,p},\Psi_{j,p}\big\ra_{L_2(\square)}=-\frac{2\pi p}{T}-\tau_0, \quad \big\la l_n(\tau_0)\Psi_{k,q},\Psi_{k,q}\big\ra_{L_2(\square)}=-\frac{2\pi q}{T}-\tau_0
\end{gather*}
Thus, the solvability conditions (\ref{3.6}) can be rewritten as
\begin{align*}
&(B(t)-K_\pm(t))c_\pm(t)=0, \quad B(t)=2t B_1-B_0(\tau_0),
\\
&c_\pm:=\begin{pmatrix}
c_1^\pm
\\
c_2^\pm
\end{pmatrix},
\quad\hphantom{(t))c_\pm(t)=0}
B_1:=\begin{pmatrix}
\frac{2\pi p}{T}+\tau_0 & 0
\\
0 & \frac{2\pi q}{T}+\tau_0
\end{pmatrix}.
\end{align*}
Therefore, $K_\pm$ are the eigenvalues of $B(t)$, and $c_\pm(t)$  are the associated eigenvectors. 
The leading terms of asymptotics (\ref{3.4}) are completely determined. 

\subsection*{Step 4: Splitting of the band functions} At this last step, on the basis of (\ref{3.4}) we analyze the behavior of the band functions as $\tau$ is close to $\tau_0$.

By an explicit analysis,
\begin{equation}
K_\pm(t)=k_1(\tau_0) t+ k_2(\tau_0) \pm \sqrt{\big(k_3(\tau_0) t+k_4(\tau_0)\big)^2+\big|b_{12}^0(\tau_0)\big|^2},\label{3.9}
\end{equation}
with $k_i$ given by  (\ref{2.18}). By (\ref{2.10}) one has $K_+(t)>K_-(t)$ for all $t\in \mathds{R}$.
Using the explicit expressions again we obtain
\begin{equation*}
2k_1(\tau_0)=-\frac{\p \L_{j,p}}{\p\tau}(\tau_0)-\frac{\p \L_{k,q}}{\p\tau}(\tau_0),\quad 2k_3(\tau_0)=\frac{\p \L_{j,p}}{\p\tau}(\tau_0)-\frac{\p \L_{k,q}}{\p\tau}(\tau_0).
\end{equation*}
In accordance with (\ref{2.9}), the derivatives in the latter formulas have opposite signs, and this shows that
\begin{equation}\label{3.15}
\big|k_1(\tau_0)\big|<\big|k_3(\tau_0)\big|\ne0,
\end{equation}
and implies the statement (\ref{A1}). Together with (\ref{3.9}) it yields
\begin{equation}\label{3.10}
\lim\limits_{|t|\to\infty} K_\pm(t)=\pm\infty.
\end{equation}
By elementary consideration one can find the minimum of  $K_+$ and the maximum of $K_-$,
\begin{gather}
\min\limits_{\mathds{R}} K_+(t)=K_+(t_+),\quad \max\limits_{\mathds{R}} K_-(t)=K_-(t_-),\nonumber
\\
t_\pm=\mp\frac{k_1(\tau_0) \big|b_{12}^0(\tau_0)\big|}{\big|k_3(\tau_0)\big|\sqrt{{\mathstrut}^{\mathstrut} k_3^2(\tau_0)
-k_1^2
(\tau_0) }}-\frac{k_4(\tau_0)}{k_3(\tau_0)},\quad
K_\pm(t_\pm)=\b_\pm(\tau_0),
\label{3.11}
\end{gather}
where $\b_\pm$ are given by (\ref{2.12}). Let us choose the parameter $a$ in (\ref{3.3}) large enough
and employ the asymptotics (\ref{3.4}), then
\begin{equation}\label{3.12}
\begin{aligned}
&\min\limits_{\tau\in[0,\pi/T]} E_+(\e,\tau)=\L_0+\e K_+(t_+)+\Odr(\e^2+\e\eta),
\\
&\max\limits_{\tau\in[0,\pi/T]} E_-(\e,\tau)=\L_0+\e K_-(t_-) +\Odr(\e^2+\e\eta),
\end{aligned}
\end{equation}
and $\max\limits_{\tau\in[0,\pi/T]} E_-(\e,\tau)<\min\limits_{\tau\in[0,\pi/T]} E_-(\e,\tau)$
for sufficiently small  $\e$ due to (\ref{3.11}).

Similar arguments are valid for $\tau\in[-\pi/T,0]$, and one arrives at the analogues of (\ref{3.4}) and (\ref{3.11}) near $-\tau_0$.
To obtain the analogues of the expressions   (\ref{3.14}), (\ref{3.9}), (\ref{3.15}), (\ref{3.10}), (\ref{3.11}) and (\ref{3.12})
one should just replace $(p,q,\tau_0)$ by $(-p,-q,-\tau_0)$ in (\ref{2.18}) and (\ref{3.14}), which gives
exactly (\ref{2.18a}) and (\ref{3.14a}). This shows that $\Op_\e$ there exists a gap  $\big(\a_l(\e),\a_r(\e)\big)$ with  $\a_{l/r}(\e)$ given by (\ref{2.11}).

The extrema of $K_\pm$ are non-degenerate, $|K_\pm(t)-K_\pm(t_\pm)|\geqslant C|t-t_\pm|^2$
with a $t$-independent constant $C$, and
\begin{equation*}
\big|E_\pm(\e,\tau_0+\e t)-\L_0-\e K_\pm(t_\pm)\big| \geqslant C\e|t-t_\pm|^2+\Odr(\e^2+\e\eta).
\end{equation*}
This shows that the maximum points of $E_-(\e,\cdot)$ and the minimum points of  $E_+(\e,\cdot)$ on $\tau\in[0,\pi/T]$
are separated from $\tau_0+\e t_+$ and $\tau_0+\e t_-$ by a distance of at most $\Odr(\e^{3/2}+\e\eta^{1/2})$,
which proves the estimates (\ref{2.13}) and (\ref{2.14}) and completes the proof of the statement (\ref{A2}).

To prove (\ref{A3}) we remark first that  $\Psi_{j,p}=\overline{\Psi_{j,-p}}$. If all the coefficients of the operator  $\Op_\e$ are real-valued, then it commutes with the complex conjugation. In particular, $\mathcal{L}_0(-\tau)\Psi_{j,-p}=\overline{\mathcal{L}_0(\tau)\Psi_{j,p}}$, and  $\beta_\pm(\tau_0)=\beta_\pm(-\tau_0)$. Now the validity of \eqref{2.16} follows trivially from
(\ref{A1}).

In the case $\tau_0=0$ the condition (\ref{2.16}) is reduced to the strict inequality $\b_-(0)<\b_+(0)$,
which is true due to (\ref{A1}). Theorem~\ref{th2.1} is proved.

\section{Examples}

Let us discuss in greater detail several specific situations
to which one can apply Theorem~\ref{th2.1}.
We will focus mainly on checking the conditions (\ref{2.10}) and (\ref{2.16}).

\subsection{Perturbation by a potential}

Adding a  real-valued periodic potential may be viewed as one of the simplest examples.
Consider the operators $\Op_\e:=\Op_0+\e V$, where $V:\Om\to \RR$ is a bounded
measurable function with $V(x',x_n+T)= V(x',x_n)$ for all $(x',x_n)\in \Om$.
This is indeed covered by the initial construction with $\mathcal{L}_\e=A_0\equiv V$.
In accordance with the conclusion (\ref{A3})
it is sufficient to check the condition (\ref{2.10}), which loses the dependence on $\tau_0$ and looks very simple,
\begin{align*}
\big\la V\Psi_{j,p},\Psi_{k,q}\big\ra_{L_2(\square)}\equiv \int\limits_{\square} V(x)\psi_j(x')\psi_k(x')
\E^{2\pi \iu (p-q)x_n/T}\di x\not=0,
\end{align*}
%This inequality is satisfied by a large class of potentials, %and 
%KP
which can be interpreted as a non-vanishing condition of a certain Fourier coefficient.
%This inequality means that the Fourier coefficient at $\psi_j(x')\psi_k(x')\E^{2\pi \iu (p-q)x_n/T}$ is non-zero.
This is a very weak condition and it describes a very large class of potentials. Hence, 
we can conclude that
a generic potential perturbation  opens the described gap in the spectrum
as far as the cross-section and the period obey the relations discussed in Remark \ref{rm2.2}.

%KP

\subsection{Perturbation by a magnetic field}

Another natural perturbation is the action of a weak magnetic field. The perturbed operator has the form
$\Op_\e=(\iu\nabla+\e A)^2$, where $A=(A_1,\ldots,A_n)$ is magnetic vector potential,
and the perturbing operator is given by
\begin{equation*}
\mathcal{L}_\e:=\iu \sum\limits_{i=1}^{n} \left( A_i \frac{\p}{\p x_i}+\frac{\p}{\p x_i} A_i \right)+\e|A|^2.
\end{equation*}
This results in
\begin{equation*}
\mathcal{L}_0(\tau)=\iu\sum\limits_{i=1}^{n} \left(A_i\frac{\p}{\p x_i} + \frac{\p}{\p x_i} A_i\right)-2\tau A_n.
\end{equation*}
For $\tau\geqslant0$ we calculate the entries of the matrix $B_0(\tau)$ using the integration by parts
\begin{equation}\label{4.4}
\begin{aligned}
b_{12}^0(\tau)=&\overline{b_{21}^0}(\tau)= \iu \sum\limits_{i=1}^{n} \left( \left\la A_i \frac{\p\Psi_{j,p}}{\p x_i}, \Psi_{k,q}\right\ra_{L_2(\square)}- \left\la A_i \Psi_{j,p}, \frac{\p\Psi_{k,q}}{\p x_i}\right\ra_{L_2(\square)}
\right)
\\
&-2\tau\la A_n \Psi_{j,p}, \Psi_{k,q}\ra_{L_2(\square)}
\\
=&\frac{\iu}{T} \sum\limits_{i=1}^{n-1} \int\limits_{\square} \E^{\frac{2\pi\iu(p-q)}{T}x_n} A_i(x) \left(\frac{\p\psi_j}{\p x_i}(x')\psi_k(x')- \frac{\p\psi_k}{\p x_i}(x')\psi_j(x')  \right)\di x
\\
& -\frac{2}{T} \left(\frac{\pi(p+q)}{T}+\tau\right) \int\limits_{\square} \E^{\frac{2\pi\iu (p-q)}{T}x_n} A_n(x)\psi_j(x')\psi_k(x')\di x,
\\
b_{11}^0(\tau)=&
-\frac{2}{T} \left(\frac{2\pi p}{T}+\tau\right) \int\limits_{\square}  A_n(x)\psi_j^2(x')\di x,
\\
b_{22}^0(\tau)=&
-\frac{2}{T} \left(\frac{2\pi q}{T}+\tau\right) \int\limits_{\square}  A_n(x)\psi_k^2(x')\di x.
\end{aligned}
\end{equation}
Similar calculations for $\tau<0$ give the relations
\begin{equation}\label{4.6}
b_{im}^0(-\tau)=-\overline{b_{mi}^0(\tau)},\quad i,m=1,2, \qquad k_i(-\tau)=-k_i(\tau),\quad i=2,4,
\end{equation}
and
\begin{equation}\label{4.5}
\b_\pm(-\tau)=-\b_\mp(\tau).
\end{equation}
Taking into account (\ref{A1}) one can see that the condition (\ref{2.16}) becomes equivalent to
$\b_+(\tau_0)>\b_-(-\tau_0)$ and $\b_+(-\tau_0)>\b_-(\tau_0)$.
By (\ref{4.5}) this reduces to
\begin{equation}\label{4.7a}
\pm\b_\pm(\tau_0)>0,
\end{equation}
which is an equivalent compact form for (\ref{2.16}).

Since $p$ and $q$ must have opposite signs  (see Remark~\ref{rm2.2}), one may assume without loss of generality $p\geqslant 0$ and $q<0$, then
the function $k_3(\tau_0)$ becomes positive. Substituting (\ref{2.12}), (\ref{2.18}) and (\ref{4.4}) into (\ref{4.7a}) one arrives at
\begin{gather}
\label{4.7}
\begin{aligned}
|b_{12}^0(\tau_0)|\sqrt{k_3^2(\tau_0)-k_1^2(\tau_0)}&>|k_1(\tau_0)k_4(\tau_0) -k_3(\tau_0)k_2(\tau_0)|,
\\
|b_{12}^0(\tau_0)|\sqrt{k_3^2(\tau_0)-k_1^2(\tau_0)}&>
\left|a_{11} \left(\frac{2\pi p}{T}+\tau_0\right)^2- a_{22} \left(\frac{2\pi q}{T}+\tau_0\right)^2\right|,
\end{aligned}
\\
\text{where }
a_{11}:=\frac{1}{T}\int\limits_{\square} A_n(x)\psi_j^2(x')\di x \text{ and } a_{22}:=\frac{1}{T}\int\limits_{\square} A_n(x)\psi_k^2(x')\di x.\label{4.10}
\end{gather}
These inequalities together with
\begin{equation}\label{4.8}
b_{12}^0(\tau_0)\not=0
\end{equation}
give the final set of sufficient conditions guaranteeing the existence of a gap for $\Op_\e$.

Let us now give an example of a specific magnetic field satisfying (\ref{4.7}) and (\ref{4.8}).
We restrict our attention to the functions $A_n(x)$ obeying
\begin{equation*}
\int\limits_{0}^{T}A_n(x)\di x_n=0\quad\text{for all}\quad x'\in\om,
\end{equation*}
then (\ref{4.10}) shows that $a_{11}=a_{22}=0$,
and the inequality (\ref{4.7}) is satisfied once the condition (\ref{4.8}) is valid.
Assume additionally $A_1(x)\equiv A_2(x)\equiv \ldots \equiv A_{n-1}(x)\equiv 0$,
then the expression (\ref{4.4}) for $b_{12}^0$ can be considerably simplified, and Eq. (\ref{4.8})
becomes equivalent to
\begin{equation*}
 \left(\frac{\pi(p+q)}{T}+\tau_0\right) \int\limits_{\om}\di x' \psi_k(x')\psi_j(x')\int\limits_{0}^{T}\di x_n A_n(x)
\E^{\frac{2\pi\iu(p-q)}{T}x_n}\not=0.
\end{equation*}
For $\tau_0=0$ and $\tau_0=\pi/T$ the coefficient before the integral vanishes, see Remark~\ref{rm2.2},
and our constructions do not allow to identify the gap opening (if a gap opens, its length is of order $o(\e)$).
On the other hand, for $\tau_0 \in(0,\pi/T)$ the coefficient is non-zero,
and all the required conditions are satisfied if one takes
\begin{equation*}
A_n(x)= \psi_k(x')\psi_j(x')\vp_1(x')\cos\frac{2\pi\iu(p-q)}{T} \,x_n+\vp_2(x),
\end{equation*}
where $\vp_1\in C^1(\overline{\om})$ is an arbitrary positive function and $\vp_2\in C^1(\overline{\Om})$ is an arbitrary $T$-periodic w.r.t. $x_n$ function satisfying the condition
\begin{equation*}
\int\limits_{0}^{T} \vp_2(x)\di x_n=\int\limits_{0}^{T} \E^{\frac{2\pi(p-q)}{T}x_n}\vp_2(x)\di x_n=0\quad \quad\text{for all}\quad x'\in\om.
\end{equation*}
For instance, one can %
take simply $\vp_1\equiv 1$ and $\vp_2\equiv 0$. Thus we can conclude that the magnetic field can open a gap
corresponding to the values of the quasi-momentum different from $0$ and $\pm\pi/T$.

\subsection{Deformation of boundary}

\begin{figure}
\begin{center}\includegraphics[height=60mm]{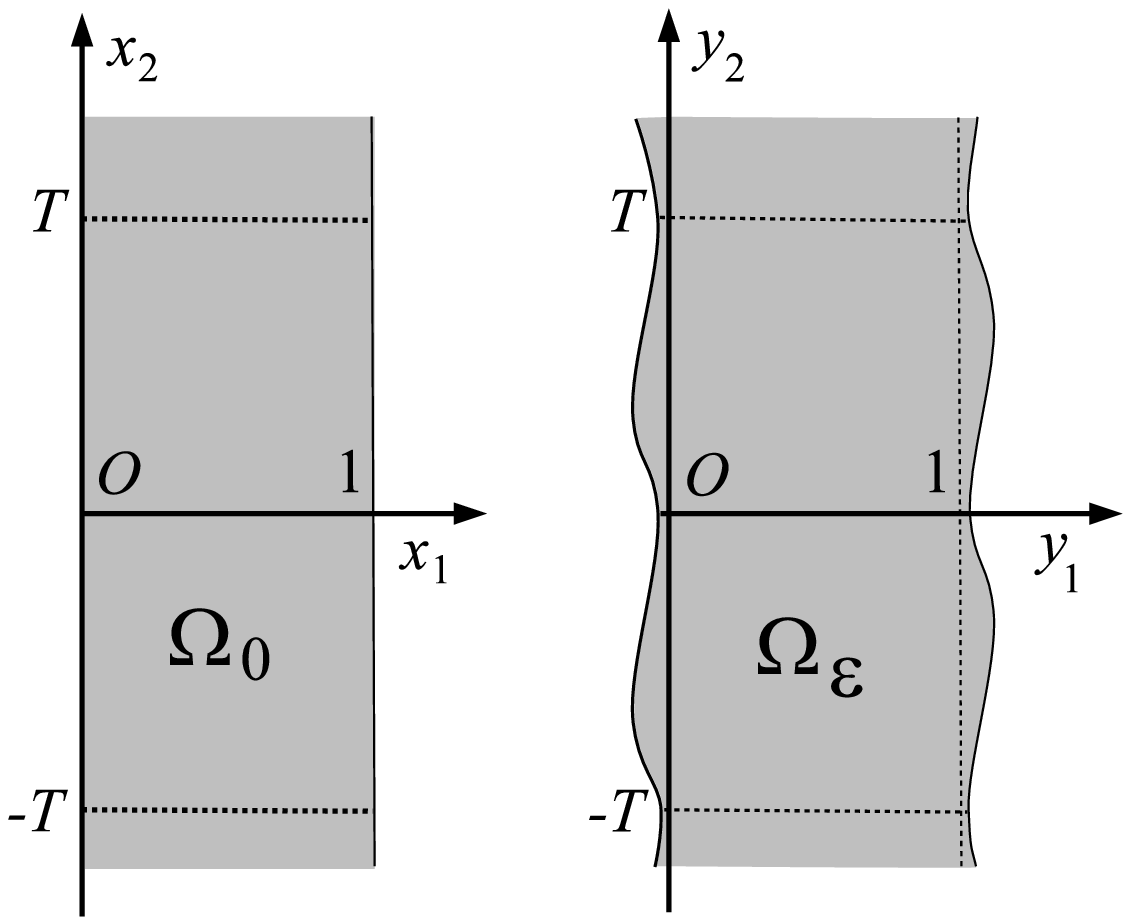}
\caption{The domains $\Om_0$ and its image $\Om_\e=\Phi_\e(\Om_0)$.\label{fig-def}}
\end{center}
\end{figure}

Let us study a geometric perturbation consisting a periodic deformation of the boundary. We consider the two-dimensional strip
\begin{equation*}
\Om_\e:=\{(y_1,y_2): \e h_-(y_2)<y_1<1+\e h_+(y_2)\},\quad y=(y_1,y_2),
\end{equation*}
where $h_\pm$ are smooth $T$-periodic functions. Denote by  $\widetilde{\Op}_\e$ the positive Dirichlet Laplacian in $\Om_\e$. Let us show
that such operators are covered by the aforementioned construction. Let us denote $h:=h_+-h_-$.
For small $\e$ the map $\Phi:\mathds{R}^2\to \mathds{R}^2$,
\[
\Phi_\e(x_1,x_2)=\begin{pmatrix}
\e h_-(x_2)+\big(1+ \e h(x_2)\big)\, x_1\\
x_2
\end{pmatrix}
\]
is a diffeomorphism, and $\Om_\e=\Phi_\e(\Om_0)$, see Figure \ref{fig-def}.
Let $J\Phi_\e$ be the Jacobi matrix of $\Phi_\e$, then the map
\begin{equation}\label{eq-ue}
U_\e:L_2(\Om_\e)\to L_2(\Om_0), \quad U_\e f= \sqrt{\det J\Phi_\e}\, f\circ \Phi_\e,
\end{equation}
is unitary, and the operator $\Op_\e:=U_\e \widetilde \Op_\e U_\e^*$
corresponds to the sesquilinear form
\begin{equation}\label{eq-form}
q_\e(u,v)=
\int_{\Om_0} \nabla \Big( \dfrac{u(x)}{\sqrt{\det J\Phi_\e(x)}}\Big)
\cdot (J\Phi_\e\mathstrut^t \cdot J\Phi_\e)^{-1}\nabla \Big( \dfrac{\overline{v(x)}}{\sqrt{\det J\Phi_\e(x)}}\Big)
 \det J\Phi_\e(x)\, \di x,
\end{equation}
i.e.,
\[
\Op_\e=-\dfrac{1}{\det J\Phi_\e}\,\text{div} \bigg(
 \det J\Phi_\e (J\Phi_\e^t \cdot J\Phi_\e)^{-1}
\nabla  \dfrac{1}{\sqrt{\det J\Phi_\e}}
\bigg).
\]
In our case
\[
J\Phi_\e (x_1,x_2)=\begin{pmatrix}
1+\e h(x_2) & \e\big( h_-'(x_2)+x_1h'(x_2)\big)\\
0 & 1
\end{pmatrix},
\]
which shows that $\Op_\e$ has real-valued coefficients and has the required representation $\Op_\e=\Op_0+\e\mathcal{L}_\e$.
By direct calculation we get
\begin{multline*}
\langle\mathcal{L}_0 u, v\rangle_{L_2(\RR^2)}=
-\int_{\Omega_0}
\dfrac{h'(x_2)}{2}\Big[
u(x) \overline{\dfrac{\partial  v(x)}{\partial x_2}}
+
\dfrac{\partial u (x)}{\partial x_2}\, \overline{v(x)}
\Big]+2 h(x_2)
\dfrac{\partial u(x)}{\partial x_1}
\overline{\dfrac{\partial v(x)}{\partial x_1}}\\
+\big(h'_-(x_2) +x_1 h'(x_2)\big)
\Big(
\dfrac{\partial u(x)}{\partial x_1}
\overline{\dfrac{\partial v(x)}{\partial x_2}}
+\dfrac{\partial u(x)}{\partial x_2}
\overline{\dfrac{\partial  v(x)}{\partial x_1}}
\Big) \di x.
\end{multline*}
Let us find a class of functions $h_\pm$ for which the matrix entry   \eqref{2.10} is non-zero. The transversal orthonormal
eigenfunctions $\psi_j$ are given by the explicit expressions $\psi_j(x)=\sqrt{2}\sin(\pi j x)$, $j=1,2,\dots$.
Employing the notation
\[
\Hat h_m=\int_0^{T}\E^{2\pi \iu m x_2/T} h(x_2)\di x_2,
\]
by direct calculations we obtain
\[
I:= \big\langle \mathcal{L}_0(\tau_0) \Psi_{j,p},\Psi_{k,q}\big\rangle_{L_2(\square)}
=\big\langle \mathcal{L}_0 \Psi_{j,p}\E^{\iu \tau_0 x_2},\Psi_{k,q}\E^{\iu \tau_0 x_2}\big\rangle_{L_2(\square)}
=-\dfrac{I_1+I_2+I_3}{T},
\]
where
\begin{align*}
I_1&=\dfrac{1}{2}\int\limits_\square \dfrac{2\pi\iu(p-q)}{T}h'(x_2)\E^{2\pi \iu (p-q)x_2/T}\psi_j(x_1)\psi_k(x_1)\di x
=\dfrac{2\pi^2(p-q)^2}{T^2} \Hat h_{p-q}\,\delta_{jk},\\
I_2&=2\int\limits_\square h(x_2)\E^{2\pi \iu (p-q)x_2/T}\psi'_j(x_1)\psi'_k(x_1)\di x=2(\pi j)^2 \Hat h_{p-q}\,\delta_{jk},\\
I_3&=\iu\int\limits_\square \big( h'_-(x_2)+x_1 h'(x_2)\big) \E^{2\pi \iu (p-q)x_2/T}\\
&\quad\cdot
\Big(
\big(\tau_0+\dfrac{2\pi p}{T}\big)\psi_j(x_1)\psi'_k(x_1)
-
\big(\tau_0+\dfrac{2\pi q}{T}\big)\psi'_j(x_1)\psi_k(x_1)
\Big)\di x.
\end{align*}
We observe that for $j,k\in\NN$ one has
\begin{align*}
\int_0^1 \psi_j(x_1) \psi'_k(x_1)\di x_1&=\begin{cases}
0, & j=k,\\
\big(1+(-1)^{j+k+1}\big)\,\dfrac{2jk}{j^2-k^2},&j\ne k,
\end{cases}\\
\int_0^1 x_1\psi_j(x_1) \psi'_k(x_1)\di x_1&=\begin{cases}
-\dfrac{1}{2}, %
 & j=k,\\
(-1)^{j+k+1} \, \dfrac{2jk}{j^2-k^2}, & j\ne k.
\end{cases}
\end{align*}

For the central intersection, $\tau_0=0$, we have $q=-p$, $j=k=1$, and
\[
I_3=-\dfrac{2\pi \iu p}{T} \int_0^T h'(x_2)\E^{4\pi \iu p x_2/T}dx_1=
-\dfrac{8\pi^2p^2}{T^2}\Hat h_{2p},
\]
which gives $I=- \dfrac{2\pi^2}{T}\,\Hat h_{2p}$.
Thus, the condition $\Hat h_{2p}\ne 0$ is sufficient to satisfy  \eqref{2.10}.

For $\tau_0=\pi/T$ we have $j=k=1$ and $q=-p-1$ and
$I=- \dfrac{2\pi^2}{T}\,\Hat h_{2p+1}$,
so the condition \eqref{2.10} holds true for $\widehat h_{2p+1}\ne 0$.

Let us now consider the non-symmetric case  $\tau_0\in (0, \pi/T)$. Without loss of generality we let $j=1$ and  $k=2$,
and one has immediately $I_1=I_2=0$. On the other hand, by using \eqref{eq-l12} we get
\begin{align*}
I_3&=-\dfrac{2\iu}{3} \Big( 2\tau_0 + \dfrac{2\pi(p+q)}{T}\Big)
\int_0^T
\Big(
2h'_-(x_2)+ h'(x_2)
\Big)
\E^{2\pi \iu (p-q)x_2/T}dx_2\\
&=
\dfrac{2\iu}{3} \Big( 2\tau_0 + \dfrac{ 2\pi(p+q)}{T}\Big)
\int_0^T
\Big(
h'_-(x_2)+ h'_+(x_2)
\Big)
\E^{2\pi \iu (p-q)x_2/T}dx_2\\
&=\dfrac{2}{3}\dfrac{2\pi (p-q)}{T}\Big(\dfrac{2\pi(p+q)}{T}+2\tau_0\Big)
\widehat{(h_-+h_+)}_{p-q}\\
&=2\pi^2 \widehat{(h_-+h_+)}_{p-q}.
\end{align*}
Thus, the condition \eqref{2.10} holds true as $\widehat{(h_-+h_+)}_{p-q}\ne 0$.

It is an interesting fact that for $\tau_0=0$ and $\tau_0=\pi/T$ the sufficient conditions
for a gap opening are formulated in terms of the Fourier coefficients of the function $h:=h_+-h_-$, while for $\tau_0\in (0,\pi/T)$ the same role is played by the function $h_++h_-$. In other words, for $\tau_0=0$ and $\tau_0=\pi/T$ the gap opening is controlled, at the first order,
by the strip width, while for $\tau\in(0,\pi/T)$ the same role is played by the sum of the side variations.

\subsection{Three-dimensional rod with a periodic twisting}

\begin{figure}
\centering

\begin{tabular}{ccc}
\includegraphics[height=60mm]{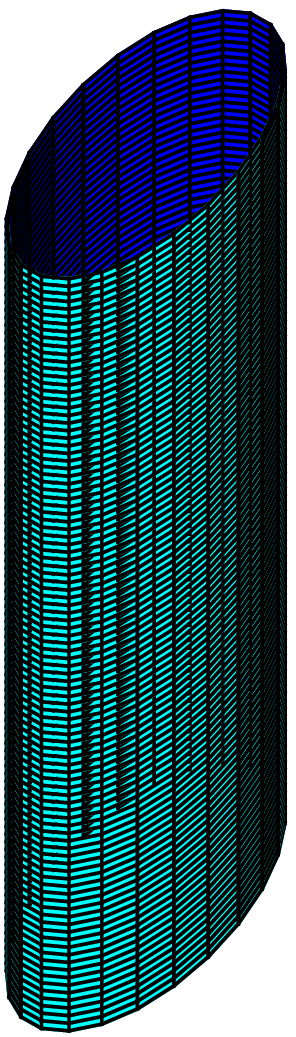}
 & \qquad\qquad&
\includegraphics[height=60mm]{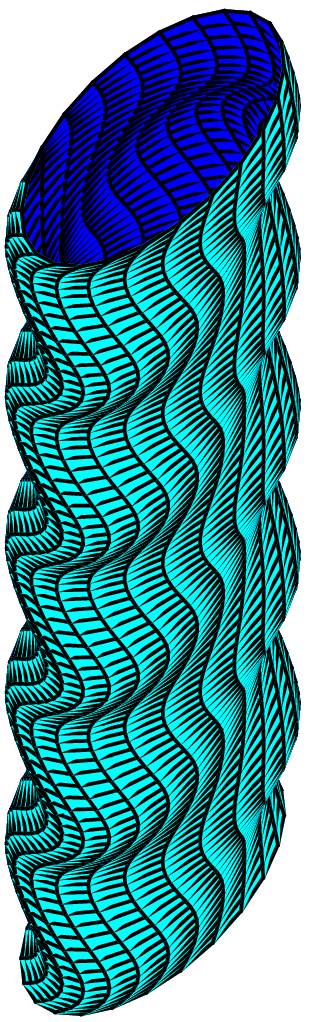}
\end{tabular}
\caption{The surface of a straight cylinder $\Om_0$ and the surface of an associated twisted cylinder $\Om_\e$ \label{fig5}}
\end{figure}

Let $\theta:\RR\to\RR$ be a smooth $T$-periodic function. For  $\e>0$ consider a diffeomorpisim $\Phi_\e:\RR^3\to\RR^3$,
\[
\Phi_\e(x_1,x_2,x_3)=\begin{pmatrix}
\cos \big(\e\theta(x_3)\big) x_1 - \sin\big(\e\theta(x_3)\big) x_2\\
\sin \big(\e\theta(x_3)\big) x_1 + \cos \big( \e\theta(x_3)\big) x_2\\
x_3
\end{pmatrix},
\]
and denote by $\om$ we denote a two-dimensional connected domain with a piece-smooth Lipshitz boundary.
Let $\Omega_0:=\omega\times\RR$ and $\Omega_\e:=\Phi_\e(\Omega_0)$.
For $\e=0$ we just get a straight cylinder with a constant cross-section $\om$, while
for $\e\ne 0$
the cross-section is rotated around the axis $Oy_3$ by the angle $\e\theta(y_3)$ w.r.t. the initial position
in each plane $y_3=\text{const}$, see Fig.~\ref{fig5}.

Let us denote by $\widetilde \Op_\e$ the Dirichlet Laplacian in $\Omega_\e$. By analogy with \eqref{eq-form} one can show that this geometric perturbation can be reduced to an additive perturbation.  Since these constructions were already discussed in various contexts, we employ the ready expression for the perturbing operator obtained in \cite[Eq. (15)]{EKK}.
Denote $\Op_\e:=U_\e \widetilde \Op_\e U_\e^*$ with the unitary
$U_\e$ defined as in  \eqref{eq-ue}, then, for any $u,v \in \Ho^2(\Om)$,
\begin{align*}
\langle  \Op_\e u,v\rangle_{L_2(\Om)}&=\int\limits_{\Om} \nabla u(x) \cdot G_\e(x) \nabla \overline{v(x)} \di x\\
\text{with } G_\e(x)&=\begin{pmatrix}
1 + \big(\e x_2 \theta'(x_3)\big)^2 & -\e^2 x_1 x_2 \big(\theta'(x_3)\big)^2 & \e x_2 \theta'(x_3)\\
-\e^2 x_1 x_2 \big(\theta'(x_3)\big)^2& 1 + \big(\e x_1 \theta'(x_3)\big)^2 & -\e x_1 \theta'(x_3)\\
\e x_2 \theta'(x_3) & -\e x_1 \theta'(x_3) & 0
\end{pmatrix}.
\end{align*}
Thus, $\Op_\e=-\Delta + \e \mathcal{L}_\e$, where the perturbation $\mathcal{L}_\e$ satisfies all the
required conditions, and its leading term $\mathcal{L}_0$ is given by
\begin{align*}
\langle \mathcal{L}_0 u,v\rangle_{L_2(\Om)}=&\int\limits_\Omega
\theta'(x_3)x_2
\Big(
\dfrac{\partial u}{\partial x_1}(x)\overline{\dfrac{\partial v}{\partial x_3}}(x)
+
\dfrac{\partial u}{\partial x_3}\overline{\dfrac{\partial v}{\partial x_1}}
\Big)(x)\\
&-
\theta'(x_3)x_1
\Big(
\dfrac{\partial u}{\partial x_2}(x)\overline{\dfrac{\partial  v}{\partial x_3}}(x)
+
\dfrac{\partial u}{\partial x_3}(x)\overline{\dfrac{\partial  v}{\partial x_2}}(x)
\Big)
\di x.
\end{align*}
We note that $\Op_\e$ commutes with the complex conjugation, and it remains to
find the conditions guaranteeing the validity of \eqref{2.10}. There holds
\begin{gather*}
I:= \langle \mathcal{L}_0(\tau_0) \Psi_{j,p},\Psi_{k,q}\rangle_{L_2(\square)}
=\langle \mathcal{L}_0 \Psi_{j,p}\E^{\iu \tau_0 x_3},\Psi_{k,q}\E^{\iu \tau_0 x_3}\rangle_{L_2(\square)}\\
=
\dfrac{1}{T} \int_\Omega
\theta'(x_3)x_2 \E^{2\pi\iu(p-q)x_3/T}\\
\cdot\left[
-
\iu\left(\dfrac{2\pi q}{T}+\tau_0\right)\dfrac{\partial \psi_j}{\partial x_1}(x_1,x_2)\psi_k(x_1,x_2)+
\iu\left(\dfrac{2\pi p}{T}+\tau_0\right)\psi_j(x_1,x_2)\dfrac{\partial \psi_k}{\partial x_1}(x_1,x_2)\right]
\\
-
\theta'(x_3)x_1 \E^{2\pi\iu(p-q)x_3/T}\\
\cdot\left[
-
\iu\left(\dfrac{2\pi q}{T}+\tau_0\right)\dfrac{\partial \psi_j}{\partial x_2}(x_1,x_2)\psi_k(x_1,x_2)+
\iu\left(\dfrac{2\pi p}{T}+\tau_0\right)\psi_j(x_1,x_2)\dfrac{\partial \psi_k}{\partial x_2}(x_1,x_2)\right]
\mathrm{d} x\\
= a(b-c),
\end{gather*}
where
\begin{align*}
a&=\dfrac{\iu}{T}\int_{0}^{T}
\theta'(x_3)\E^{2\pi\iu(p-q)x_3/T}\di x_3=
\dfrac{2\pi(p-q)}{T^2}\int_{0}^{T}
\theta(x_3)\E^{2\pi\iu(p-q)x_3/T}\di x_3,\\
b&=\iu\Big(\dfrac{2\pi p}{T}+\tau_0\Big)
\int\limits_\om
x_2 \Big(\psi_j \dfrac{\partial \psi_k}{\partial x_1}+\dfrac{\partial \psi_j}{\partial x_1}\psi _k\Big)\di x_1 \di x_2\\
&\quad{}- \iu(\dfrac{2\pi p}{T}+\tau_0)
\int\limits_\om
x_1 \Big(\psi_j \dfrac{\partial \psi_k}{\partial x_2}+\dfrac{\partial \psi_j}{\partial x_2}\psi _k\Big)\di x_1 \di x_2,\\
c&= \Big(\dfrac{2\pi(p+q)}{T}+2\tau_0\Big)
\int_\om
\Big(
x_2 \dfrac{\partial \psi_j}{\partial x_1}\psi_k
-
x_1 \dfrac{\partial \psi_j}{\partial x_2}\psi_k
\Big)\di x_1 \di x_2.
\end{align*}
One observes that the coefficient $a$ does not depend on the cross-section and can be chosen non-zero by an appropriate choice of the twisting function  $\theta$. The Green's formula gives the identities
\begin{align*}
\int_\om
x_2 \Big(\psi_j \dfrac{\partial \psi_k}{\partial x_1}+\dfrac{\partial \psi_j}{\partial x_1}\psi _k\Big)\di x_1 \di x_2&=
\int\limits_\om
x_2 \dfrac\partial{\partial x_1} (\psi_j \psi_k)\di x_1 \di x_2
=\int_{\partial \omega} x_2 \psi_j \psi_k dx_2,\\
\int\limits_\om
x_1 \Big(\psi_j \dfrac{\partial \psi_k}{\partial x_2}+\dfrac{\partial \psi_j}{\partial x_2}\psi _k\Big)\di x_1 \di x_2&=
\int\limits_\om
x_1 \dfrac\partial{\partial x_2} (\psi_j \psi_k)\di x_1 \di x_2=-\int\limits_{\partial \omega} x_1 \psi_j \psi_k \di x_1.
\end{align*}
Since the both functions  $\psi_j$ and $\psi_k$ vanish at the boundary $\partial\omega$, one has
\[
\int_{\partial \omega} x_1 \psi_j \psi_k \di x_1=\int_{\partial \omega} x_2 \psi_j \psi_k \di x_2=0,
\]
which gives $b\equiv 0$.

Let us study the remaining coefficient $c$. In the case of a symmetric intersection of the unperturbed band functions, i.e., for   $\tau_0=0$ or $\tau=\pi/T$, we have $j=k$ and $p\ne q$, which implies by \eqref{eq-l12} that $c=0$ and $I=0$.
Therefore, the assumptions of Theorem~\ref{th2.1} are not satisfied,
and the first order perturbation theory does not allow us to identify a gap opening
for quasi-momenta close to the center and to the end points of the Brillouin zone.

For $\tau_0\notin(0,\pi/T)$ we have $j\ne k$ and $p\ne q$, and by  \eqref{eq-l12} we conclude that $2\pi(p+q)/T+2\tau_0\ne 0$.
Let us show that the integral entering the expression for $c$ does not vanish at least for some specific domains $\om$.
Without loss of generality we let  $j=1$ and $k=2$, and take as an example $\om=(0,\pi)\times(0,\pi/\alpha)$ with
$\alpha>1$. Denoting $C=2\sqrt \alpha/\pi$, we have
\begin{align*}
\lambda_1&=1+\alpha^2, & \psi_1(x_1,x_2)&=C \sin (x_1) \sin (\alpha x_2),\\
\lambda_2&=4+\alpha^2, & \psi_2(x_1,x_2)&=C\sin (2x_1) \sin(\alpha x_2),
\end{align*}
and
\[
\dfrac{\partial \psi_1}{\partial x_1}(x_1,x_2)=C \cos(x_1)\sin(\alpha x_2),\quad
\dfrac{\partial \psi_1}{\partial x_2}(x_1,x_2)=\alpha C \sin(x_1)\cos(\alpha x_2).
\]
It yields
\begin{align*}
x_2\dfrac{\partial \psi_1}{\partial x_1}\psi_2&=\dfrac{C^2}{4}\, x_2  \big( \sin(3x_1)+\sin(x_1)\big)\cdot\big(1-\cos(2\alpha x_2)\big),\\
x_1\dfrac{\partial \psi_1}{\partial x_2}\psi_2&=\dfrac{\alpha C^2}{4}\, x_1 \big(\cos (x_1)-\cos(3x_1)\big)\cdot \sin(2\alpha x_2),
\end{align*}
and
\begin{align*}
\int\limits_\om x_2\dfrac{\partial \psi_1}{\partial x_1}\psi_2 \di x_1 \di x_2
&= \dfrac{C^2}{4}
\int_0^\pi \big( \sin(3x_1)+\sin(x_1)\big) \di x_1\\
&\quad \cdot \int_0^{\pi/\alpha}x_2 \big(1-\cos(2\alpha x_2)\big)\di x_2
=\dfrac{\alpha}{\pi^2}\dfrac{8}{3} \dfrac{\pi^2}{2\alpha^2}=\dfrac{4}{3\alpha},\\
\int\limits_\om x_1\dfrac{\partial \psi_1}{\partial x_2}\psi_2 \di x_1 \di x_2&=\dfrac{\alpha C^2}{4}\,
\int\limits_0^\pi x_1 \big(\cos (x_1)-\cos(3x_1)\big)\di x_1
\int\limits_0^{\pi/\alpha} \sin(2\alpha x_2)\di x_2=0.
\end{align*}
Finally, by employing \eqref{eq-l12} we obtain
\begin{align*}
I&=-ac=-\Big( \dfrac{2\pi(p+q)}{T}+2\tau_0\Big) \, \dfrac{4}{3\alpha} \dfrac{2\pi(p-q)}{T^2}\int_{0}^{T}
\theta(x_3)\E^{2\pi\iu(p-q)x_3/T}\di x_3\\
&=-\dfrac{4\pi^2}{\alpha T}\int_{0}^{T}\theta(x_3)\E^{2\pi\iu(p-q)x_3/T}\di x_3,
\end{align*}
and $I\ne 0$, if the Fourier coefficient in the previous expression is non-zero.

\section*{Acknowledgments}

D.B. was partially supported by RFBR (grant no. 13-01-91052-NCNI-a) and Federal Targeted Program (agreement no. 14.B37.21.0358).
K.P. was partially supported by ANR NOSEVOL, GDR Dynamique quantique and PICS CNRS-RFBR ASAPO.


\begin{thebibliography}{999}


\bibitem{kuch}
P. Kuchment: \emph{Floquet theory for partial differential equations.} Birkh\"auser Verlag, Basel, 1993


\bibitem{PK} P. Kuchment: \emph{The Mathematics of Photonic Crystals.}
In: G. Bao, L. Cowsar, W.~Masters (Eds): \emph{Mathematical modeling in optical science}
(Frontiers in Applied Mathematics, vol. 22), pp.~207--272,  SIAM, Philadelphia (2001).


\bibitem{HP} R. Hempel, O. Post: \emph{Spectral gaps for periodic elliptic operators with high contrast: an overview.}
In: H. G. W. Begehr, R. P. Gilbert, M. W. Wong (Eds):
\emph{Progress in Analysis. Proceedings of the 3rd International ISAAC Congress Berlin 2001. Vol. 1},
pp.~577--587, World Scientific Publishing Co., Inc., River Edge, NJ (2003).


\bibitem{LP} F. Lledo, O. Post: \emph{Generating spectral gaps by geometry.}
In: J. C. Mour\~ao et al. (Eds): \emph{Prospects in mathematical physics}, pp.~~159--169 (Contemp. Math., vol.~437),
Amer. Math. Soc., Providence, RI, 2007.


\bibitem{cnp}
G. Cardone, S. A. Nazarov, C. Perugia:
\emph{A gap in the essential spectrum of a cylindrical waveguide with a periodic perturbation of the surface.}
Math. Nachr. {\bf 283} (2010) 1222--1244.

\bibitem{naz2}  S. A. Nazarov: \emph{A Gap in the Essential Spectrum of the Neumann Problem for an Elliptic System in a Periodic Domain}.
Funct. Anal. Appl., 43:3 (2009), 239-241.

\bibitem{naz1} S. A. Nazarov: \emph{Opening of a Gap in the Continuous Spectrum of a Periodically Perturbed Waveguide}.  Math. Notes, 87:5 (2010), 738-756.


\bibitem{HKSW} J. Harrison, P. Kuchment, A. Sobolev, B. Winn:
\emph{On occurrence of spectral edges for periodic operators inside the Brillouin zone.}
J. Phys. A  {\bf 40} (2007) 7597--7618.


\bibitem{EKW}
P. Exner, P. Kuchment, B. Winn: \emph{On the location of spectral edges in $\ZZ$-periodic media.}
J. Phys. A {\bf 43} (2010) 474022.

\bibitem{east} M. S. P. Eastham: \emph{The spectral theory of periodic differential equations.}
Scottish Academic Press, Edinburgh, 1973.

\bibitem{acoust} S. D. M. Adams, R. V. Craster,
S. Guenneau: \emph{Bloch waves in periodic multi-layered
acoustic waveguides.}
Proc. Roy. Soc. Lond. A {\bf 464} (2008) 2669--2692.


\bibitem{bp1} D.I. Borisov, K.V. Pankrashkin:
\emph{On extrema of band functions in periodic waveguide}
Funct. Anal. Appl. (to appear).

\bibitem{bp2} D.I. Borisov,  K.V. Pankrashkin:
\emph{Gap opening and split band edges for waveguides coupled by a periodic system of small windows.} Math. Notes (to appear).

\bibitem{naz3} S. A. Nazarov: \emph{The asymptotic analysis of gaps in the spectrum of a waveguide perturbed
with a periodic family of small voids.} J. Math. Sci. (2) {\bf 186} (2012) 247--301.

\bibitem{CAMG} R. V. Craster,  T. Antonakakis,  M. Makwana,  and S. Guenneau: \emph{Dangers of using the edges of the Brillouin zone.} Phys. Rev. B. {\bf 86} (2012) 115130.

\bibitem{bent}
F. Bentosela, P. Duclos, P. Exner:
\emph{Absolute continuity in periodic thin tubes and strongly coupled leaky wires.}
Lett. Math. Phys. {\bf 65} (2003) 75--82.


\bibitem{fried}
L. Friedlander: \emph{Absolute continuity of the spectra of periodic waveguides.}
In: P. Kuchment (Ed): \emph{Waves in Periodic and Random Media:
Proceedings of an AMS-IMS-SIAM Joint Summer Research Conference on Waves in Periodic and Random Media, June 22-28, 2002, Mount Holyoke College, South Hadley} (Contemp. Math., vol. 339), pp. 37–42, AMS, Providence, RI (2003).


\bibitem{fil} I. Kachkovski\u{\i}, N. Filonov.
\emph{Absolute continuity of the Schr\"odinger operator spectrum in a multidimensional cylinder.} St. Petersburg Math. J.  {\bf 21}:1 (2010) 95--109.



\bibitem{sobol}
A. V. Sobolev, J. Walthoe:
\emph{Absolute continuity in periodic waveguides.}
Proc. London Math. Soc. (3) {\bf 85} (2002) 717--741.




\bibitem{Kato} T. Kato: \emph{Perturbation theory for linear operators.}     Springer-Verlag, Berlin, Heidelberg, New York, 1966.

\bibitem{EKK}
T. Ekholm, H. Kovarik, D. Krejcirik: \emph{A Hardy inequality in twisted waveguides.}
Arch. Ration. Mech. Anal. {\bf 188} (2008) 245--264.







\end{thebibliography}
\end{document}